\documentclass[11pt]{article}

\usepackage{fullpage,latexsym,amsmath,amsfonts}

\def\01{\{0,1\}}
\newcommand{\ceil}[1]{\lceil{#1}\rceil} 
 
\newcommand{\eps}{\varepsilon} 
\newcommand{\ket}[1]{|#1\rangle}

\newcommand{\OR}{\mbox{\rm OR}}

\newtheorem{definition}{Definition}
\newtheorem{theorem}{Theorem}

\begin{document}

\title{A note on quantum algorithms and the minimal degree of $\eps$-error polynomials for symmetric functions}
\author{
Ronald de Wolf\thanks{rdewolf@cwi.nl. Partially supported by a Veni grant from the Netherlands Organization for Scientific Research (NWO), and by the European Commission under the Integrated Project Qubit Applications (QAP) funded by the IST directorate
as Contract Number 015848.}\\
CWI Amsterdam}
\date{}
\maketitle

\begin{abstract}
The degrees of polynomials representing or approximating Boolean functions are a prominent tool
in various branches of complexity theory.  Sherstov~\cite{sherstov:inclexcl} recently characterized 
the minimal degree $deg_{\eps}(f)$ among all polynomials (over $\mathbb{R}$) 
that approximate a \emph{symmetric} function $f:\01^n\rightarrow\01$ up to worst-case error $\eps$:
$$
deg_{\eps}(f)=\widetilde{\Theta}\left(deg_{1/3}(f) + \sqrt{n\log(1/\eps)}\right).
$$
In this note we show how a tighter version (without the log-factors hidden in the $\widetilde{\Theta}$-notation), 
can be derived quite easily using the close connection between polynomials and quantum algorithms.
\end{abstract}

\section{Introduction}


Boolean functions are one of the primary objects of study in theoretical computer science.
Such functions can be represented or approximated by \emph{polynomials} in a number of ways,
and the algebraic properties of such polynomials (such as their degree) often 
give information about the complexity of the function involved.
Areas where this approach has been used include 
circuit complexity~\cite{razborov:circuit,smolensky:algebraic,beigel:poly},
complexity classes~\cite{brs:pp,beigel:perceptrons,toda:ppashardasph},
decision trees~\cite{nisan&szegedy:degree,buhrman&wolf:dectreesurvey}, 
communication complexity~\cite{buhrman&wolf:qcclower,razborov:qdisj,sherstov:qcclower,lee&shraibman:nofdisj}, 
and learning theory~\cite{mos:learningjuntas,lmmv:symmjunta}.

In this note we focus on polynomials over the field of real numbers.
An \emph{$n$-variate multilinear polynomial} $p$ is a function $p:\mathbb{R}^n\rightarrow\mathbb{R}$
that can be written as
$$
p(x_1,\ldots,x_n)=\sum_{S\subseteq[n]} a_S\prod_{i\in S}x_i,
$$
for some real numbers $a_S$.
The \emph{degree} of $p$ is $deg(p)=\min\{|S|\mid a_S\neq 0\}$.
If is well known (and easy to show) that every function $f:\01^n\rightarrow\mathbb{R}$
has a unique representation as such a polynomial; $deg(f)$ is defined as the degree 
of that polynomial.

In many applications it suffices if the polynomial is \emph{close} to $f$ instead of being equal to it:

\begin{definition}
The \emph{$\eps$-approximate degree} of $f:\01^n\rightarrow\mathbb{R}$ is
$$
deg_{\eps}(f)=\min\{deg(p)\mid \forall x\in\01^n:|p(x)-f(x)|\leq\eps\}.
$$
\end{definition}

A function $f$ is called \emph{symmetric} if its value only depends on the Hamming weight
$|x|$ of its input $x\in\01^n$. Equivalently, 
$f(x)=f(\pi(x))$ for all $x\in\01^n$ and all permutations $\pi\in S_n$.
We will restrict attention here to symmetric functions $f$.
Examples are OR, AND, PARITY, MAJORITY etc.
Since the only thing that matters is the Hamming weight $|x|$ of the input,
one can actually restrict attention to \emph{univariate} polynomials.
We say that a univariate polynomial $p$ \emph{$\eps$-approximates} 
a symmetric function $f$ if $|p(|x|)-f(x)|\leq\eps$ for all $x\in\01^n$.  
By a technique called \emph{symmetrization}~\cite{minsky&papert:perceptrons},
it turns out that for symmetric functions, the minimal degree of such univariate 
$\eps$-approximating polynomials is the same degree $deg_{\eps}(f)$ 
as for $n$-variate multilinear polynomials.
Hence we can switch back and forth between these two kinds of polynomials at will.

Paturi~\cite{paturi:degree} tightly characterized the 1/3-approximate degree $deg_{1/3}(f)$
of all symmetric $f$ (see the start of Section~\ref{secproof} for the precise statement).
Recently, Sherstov~\cite{sherstov:inclexcl} studied the dependence on the error $\eps$.
He proved the surprisingly clean result that for all $\eps\in[2^{-n},1/3]$,
$$
deg_{\eps}(f)=\widetilde{\Theta}\left(deg_{1/3}(f) + \sqrt{n\log(1/\eps)}\right),
$$
where the $\widetilde{\Theta}$ notation hides some logarithmic factors.
Note that the statement is false if $\eps\ll 2^{-n}$, since clearly $deg(f)\leq n$ for all $f$.

Sherstov gave an interesting application of his result 
in the context of the inclusion-exclusion principle of probability theory.  
Let $f:\01^n\rightarrow\01$ be a Boolean function.  
Suppose one has events $A_1,\ldots,A_n$ in some probability space, and one knows the exact 
values of $\Pr[\cap_{i\in S} A_i]$ for all sets $S\subseteq[n]$ of size at most $k$.
How well can we now estimate $\Pr[f(A_1,\ldots,A_n)]$?
Sherstov gives essentially tight bounds for this for all symmetric functions $f$,
based on his degree-result.
This generalizes earlier results for the case where $f$ is the OR 
function, i.e.~where one is estimating $\Pr[\cup_{i\in[n]}A_i]$~\cite{linial&nisan:inclexl,kls:inclexcl}.

In this note we give a different proof, for a slightly tighter version of Sherstov's degree-result:

\begin{theorem}\label{thepsdeg}
For every non-constant symmetric function $f:\01^n\rightarrow\01$ and $\eps\in[2^{-n},1/3]$:
$$
deg_{\eps}(f)=\Theta\left(deg_{1/3}(f) + \sqrt{n\log(1/\eps)}\right).
$$
\end{theorem}

Note that there are no hidden logarithmic factors anymore.
As a consequence, the result on approximate inclusion-exclusion 
is sharpened as well, but we won't elaborate on that here.

The lower bound on $deg_{\eps}(f)$ follows immediately from combining Paturi's tight bound
for $deg_{1/3}(f)$ with the tight bound on the $\eps$-approximate degree
of the OR-function proved in~\cite{bcwz:qerror}. More interestingly, our upper bound is obtained 
by exhibiting an efficient $\eps$-error \emph{quantum algorithm} for computing a symmetric function.
It is well known (at least in quantum circles) that the acceptance probability 
of a quantum algorithm that makes $T$ queries to its input can be written as 
an $n$-variate multilinear polynomial of degree at most $2T$~\cite{bbcmw:polynomialsj}.
The upper bound of Theorem~\ref{thepsdeg} actually applies to a larger class of functions, 
namely all functions that are constant when $|x|\in\{t,\ldots,n-t\}$. These functions may be 
arbitrary (possibly non-symmetric) for smaller or larger Hamming weights. 
For every such function we have $deg_{\eps}(f)=O(\sqrt{tn} + \sqrt{n\log(1/\eps)})$.

\subsection*{Discussion}

The main message of this note is that one can obtain essentially optimal polynomial 
approximations of symmetric Boolean functions by arguing about quantum algorithms.
This fits in a line of papers in recent years that prove or reprove theorems 
about various topics in classical computer science or mathematics with the help
of quantum computational techniques.  This includes results about locally decodable 
codes~\cite{kerenidis&wolf:qldcj,wehner&wolf:improvedldc},
classical proof systems for lattice problems inspired by earlier quantum proof
systems~\cite{aharonov&regev:latticeqnp,aharonov&regev:latticenpconp},
limitations on classical algorithms for local search~\cite{aaronson:localsearch}
inspired by an earlier quantum proof, a proof that the
complexity class PP is closed under intersection~\cite{aaronson:pp},
lower bounds on the rigidity of Hadamard matrices~\cite{wolf:rigidity},
classical formula size lower bounds from quantum query lower bounds~\cite{lls:advform},
and an approach to proving lower bounds for classical circuit depth
using quantum communication complexity~\cite{kerenidis:qcircuit}.

There are advantages as well as disadvantages to our approach in this note.
We feel that for someone familiar with quantum algorithms and their connection to polynomials,
our proof should be quite simple and straightforward.  Also, our bound applies to 
a larger class of functions, and is tight up to constant instead of logarithmic factors.
On the other hand, for those unfamiliar with quantum computation our proof is 
probably not that accessible. Another disadvantage is that we do not construct the 
$\eps$-approximating polynomials explicitly (though one may derive them from our quantum algorithm), 
in contrast to Sherstov's construction based on Chebyshev polynomials.

\section{Proof}\label{secproof}

Let $f:\01^n\rightarrow\01$ be a non-constant symmetric function that is constant 
if the Hamming weight $|x|$ of the input is in the interval $\{t,..,n-t\}$ 
(where $0<t\leq n/2$ is the smallest $t$ for which this holds).  
We know $deg_{1/3}(f)=\Theta(\sqrt{tn})$ from Paturi~\cite{paturi:degree}. 
In the next two subsections we provide matching upper and lower bounds on $deg_{\eps}(f)$,
thus proving Theorem~\ref{thepsdeg}.

\subsection{Upper bound on $deg_{\eps}(f)$}

Beals et al.~\cite{bbcmw:polynomialsj} showed that the acceptance probability of 
a $T$-query quantum algorithm on $n$-bit input is a multilinear $n$-variate polynomial 
$p:\mathbb{R}^n\rightarrow\mathbb{R}$ of degree at most $2T$. 
Hence it suffices to give an $\eps$-error quantum algorithm for 
$f$ that uses $O(deg_{1/3}(f) + \sqrt{n\log(1/\eps)})$ queries. The
acceptance probability of the algorithm will be our $\eps$-error polynomial.

Here is the algorithm.  It uses various quantum algorithms based on Grover's search algorithm,
which are explained in the appendix. Let $x\in\01^n$ be the input string.  
The algorithms have access to this string via \emph{queries}.  In the quantum case, one query is 
one application of the unitary that maps $\ket{i}\mapsto(-1)^{x_i}\ket{i}$.
A \emph{solution} is an index $i\in[n]$ such that $x_i=1$.
\begin{enumerate}
\item Use $t$ repeated applications of exact Grover to try to find up to $t$ solutions 
(initially assuming $|x|=t$, and ``crossing out'' in subsequent applications 
the solutions already found).  If $|x|\leq t$, then \emph{with probability 1} these repeated applications 
find all solutions.  This costs $O(\sqrt{tn})$ queries. 
\item Use $\eps/2$-error Grover to try to find one more solution. 
This costs $O(\sqrt{n\log(1/\eps)})$ queries.
\item The same as step~1, but now looking for positions of 0s instead of 1s.
\item The same as step~2, but now looking for a 0 instead of a 1.
\end{enumerate}
The total number of queries is indeed $O(\sqrt{tn} + \sqrt{n\log(1/\eps)})$.
We need to show that this gives error probability at most $\leq\eps$ for every input $x\in\01^n$.
Observe the following:
\begin{itemize}
\item if step~1 found $t$ solutions, then we know $|x|\geq t$ with probability~1 
(note that you can verify whether a given position is a solution with only 1 extra query).
\item if step~1 found fewer than $t$ solutions, but step~2 found another solution, then we
 know $|x|>t$ (for if $|x|\leq t$ then step~1 would certainly have found all solutions 
and there would be none left to be found in step~2).
\item if step~1 found fewer than $t$ solutions, but step~2 did \emph{not} find another solution, 
then the probability that there are more solutions than those found by step~1, is at most $\eps/2$
(because step~2 ran an $\eps/2$-error search algorithm which didn't find any solution).
\item similar observations for steps 3 and 4 (with 0s and 1s switching roles).
\end{itemize}
These observations imply that at the end of the 4 steps we have enough information to compute~$f$.
Note that with probability at least $1-\eps$ we can distinguish between the three cases
$|x|<t$, $|x|\in\{t,\ldots,n-t\}$, and $|x|>n-t$. 
If $|x|\in\{t,\ldots,n-t\}$ then we are done because $f$ is constant on this interval.
If $|x|<t$ then step~1 found all solutions, so we know $x$ completely and can compute $f(x)$.
If $|x|>n-t$ then step~2 found all non-solutions of $x$, and again we know $x$ completely.
In all cases we compute $f(x)$ with error probability at most $\eps$.

This algorithm even works for many non-symmetric functions: 
it suffices if $f$ is constant on all inputs with Hamming weight in $\{t,\ldots,n-t\}$; 
$f$ may be arbitrary if $|x|<t$ or $|x|>n-t$ since in these cases the algorithm actually 
determines $x$ completely, rather than just its Hamming weight.

\subsection{Lower bound on $deg_{\eps}(f)$}\label{sseclower}

We can assume $t<n/4$, because if $t\geq n/4$ then we already have a tight bound from Paturi:
$$
n\geq deg(f)\geq deg_{\eps}(f)\geq deg_{1/3}(f)=\Theta(n).
$$
Buhrman et al.~\cite{bcwz:qerror} showed for the $n$-bit OR function 
that $deg_{\eps}(\OR_n)=\Theta(\sqrt{n\log(1/\eps)})$.%
\footnote{The earlier paper by Kahn et al.~\cite{kls:inclexcl} showed a $\widetilde{\Theta}$-version of this.}
Since $t<n/4$, we can embed an OR on at least $n-2t\geq n/2$ bits into $f$ 
by fixing some of the bits to specific values.  Hence 
$$
deg_{\eps}(f)\geq \max\left(deg_{1/3}(f), \Omega(\sqrt{n\log(1/\eps)})\right)=\Omega\left(deg_{1/3}(f) + \sqrt{n\log(1/\eps)}\right).
$$

\subsection*{Acknowledgments}
Thanks to Sasha Sherstov for his paper~\cite{sherstov:inclexcl}
(which prompted this note) and some comments.

\bibliographystyle{alpha}

\newcommand{\etalchar}[1]{$^{#1}$}

\appendix

\section{Grover's algorithm and applications}


Grover's quantum algorithm~\cite{grover:search} for finding a solution (i.e.~an $i\in[n]$ such that $x_i=1$)
consists of $T$ applications of a certain unitary $G$, starting from the uniform superposition 
$\frac{1}{\sqrt{n}}\sum_{i=1}^n\ket{i}$.  We won't explain the details of $G$ here. 
Suffice it to say that each $G$ makes one quantum query, so the total number of queries is $T$.
The intuition is that $G$ changes the state by moving amplitude from non-solutions to solutions.
One can show~\cite{bhmt:countingj} that the probability that a measurement of the state 
after $T$ steps gives a solution, is exactly
$$
\left(\sin((2T+1)\theta)\right)^2, \mbox{ where }\theta=\arcsin(\sqrt{|x|/n}).
$$
If $|x|>0$ and $T=\ceil{(\pi/4)\sqrt{n/|x|}}$, then this probability is close to 1.
Hence if we know (at least approximately) the number of solutions $|x|$,
then we can find one with good probability using $O(\sqrt{n/|x|})$ queries.
If we know $|x|$ exactly, a small modification of the algorithm finds a solution \emph{with probability 1}~\cite{bhmt:countingj}.
This uses exactly $\ceil{(\pi/4)\sqrt{n/|x|}}$ queries;
we will refer to it as ``exact Grover''.

What if we don't know how many solutions there are in the input?
We can first apply Grover assuming the number of solutions is $n/2$, then assuming it is $n/4$ etc.
This finds one solution with probability at least some constant, even if we don't know 
the number of solutions.  The complexity is $\sum_{i=1}^{\log n}O(\sqrt{n/2^i})=O(\sqrt{n})$ queries.
If we know there are \emph{at least} $t$ solutions, this can be improved to $O(\sqrt{n/t})$.
We will refer to this as ``usual Grover''.

And what if we want to have probability at least $1-\eps$ of finding a solution?
Buhrman et al.~\cite{bcwz:qerror} designed an algorithm that achieves this using
$O(\sqrt{n\log(1/\eps)})$ queries, and showed 
(by proving the lower bound on $deg_{\eps}(\OR)$ mentioned in Section~\ref{sseclower})
that this complexity is optimal up to a constant factor.  Their algorithm is quite simple.
Apply exact Grover $\log(1/\eps)$ times, first assuming there is 1 solution, then assuming there
are 2 solutions, etc. If the actual number of solutions is between 1 and $\log(1/\eps)$, 
at least one solution will have been found with probability~1 by now.
If no solution has been found yet, then apply usual Grover $O(\log(1/\eps))$ many times 
assuming there are at least $t=\log(1/\eps)$ solutions. It is easy to verify that this 
has overall query complexity $O(\sqrt{n\log(1/\eps)})$ and error probability at most $\eps$.
We will refer to this as ``$\eps$-error Grover''.


De Graaf and de Wolf~\cite[Lemma~2]{graaf&wolf:qyao} observed that exact 
Grover can be used to \emph{find all solutions with probability 1}, 
as long as we know an \emph{upper bound} $t$ on the number 
of solutions. Suppose we run exact Grover $t$ times: 
the first time assuming we have exactly $t$ solutions, 
the second time assuming we have exactly $t-1$ solutions, etc.
Each time we find a solution $i$, we ``cross it out'' in the sense of modifying the input
by setting $x_i$ to 0 (this can easily be achieved by some unitary pre- and post-processing around the query).
This prevents the algorithm from finding the same solution twice.
The total number of queries used is
$$
\sum_{i=1}^t \ceil{(\pi/4)\sqrt{n/i}}\leq \frac{\pi}{2}\sqrt{tn}.
$$
To see that this finds all solutions with probability~1, observe that the assumed number of solutions
$t-i+1$ of the $i$th run always upper bounds the actual number of remaining solutions
(this ``loop invariant'' is easily proved with downward induction).
Hence if we start with at most $t$ remaining solutions, then after $t$ runs we 
end with 0 solutions---meaning all solutions have been found.

\end{document}